\documentclass[fleqn,twoside]{article}
\usepackage{espcrc2}

\usepackage{graphicx}
\usepackage[figuresright]{rotating}

\newcommand{\AmS}{{\protect\the\textfont2
  A\kern-.1667em\lower.5ex\hbox{M}\kern-.125emS}}
\title{$U_A(1)$ and spontaneous chiral symmetry breaking within a confining
model of the QCD vacuum}

\author{A.C. Kalloniatis\address
{Centre for the Subatomic Structure of Matter, 
        University of Adelaide, \\ 
        Adelaide 5000, Australia}%
        \thanks{Supported by the Australian Research Council.},
        S.N. Nedelko\address{Institute for Theoretical Physics III,
        University of Erlangen-Nuremberg, Erlangen, Germany, and BLTPh, JINR, Dubna, Russia }%
        \thanks{Supported by DFG under contract SM70/1-1 and, partially, by  BFBR grant 01-02-1720.}}
       
\begin{document}

\begin{abstract}
A model for the QCD vacuum based on domain-like background gluon fields
has previously been developed and demonstrated to exhibit confinement
of  quarks and strong correlation of the  local chirality of quark modes and 
duality of the background gluon field. On the boundary of
the j-th domain, quark fluctuations satisfy  chirality violating boundary 
conditions parametrized by a chiral angle $\alpha_j$
which is treated as a collective variable.
The dependence of the free energy of an ensemble of $N\to\infty$ domains
on $\{ \alpha_j, j=1\dots N\}$ is studied. 
For one massless flavour there are only 
two degenerate minima ($\alpha_j=0,\pi$ $\forall j$) due to the 
contribution of the anomaly
to the free energy, which suppresses the continuous $U(1)$ degeneracy. 
The residual discrete symmetry is spontaneously broken
with a quark condensate of $-(238 {\rm{MeV}})^3.$
For $N_f$ massless flavours a continuous 
$SU(N_f)_L\times SU(N_f)_R$ degeneracy appears,
while the $U_A(1)$ direction remains suppressed.
\vspace{1pc}
\end{abstract}

\maketitle

\section{INTRODUCTION}
Many of the features of 
quantum chromodynamics long expected to emerge in the
nonperturbative regime of the theory have been
realised in a range of phenomenological models
as well as demonstrated quantitatively in  
lattice gauge theory. 
Confinement of quarks,
spontaneously broken flavour chiral  symmetry 
with a nonzero
quark condensate and resolution of $U_{\rm A}(1)$
problem are believed to be intrinsically connected to each other.
Models of the QCD vacuum, which  can give insight into
what is going right with lattice simulations of QCD,
can invariably connect two of these three phenomena, but rarely all
three simultaneously. In this paper, we pursue one step further the 
exploration of the ``domain model'' for the vacuum, 
originally proposed  in \cite{NK2001}, as a scenario for
simultaneous appearance of all three phenomena:  confinement,
spontaneous chiral symmetry breaking via the appearance of
a quark condensate and a continuous
$SU(N_f)_L \times SU(N_f)_R$ degeneracy of the vacuum
for $N_f$ massless quarks, 
but without a $U_A(1)$ continuous
degeneracy of ground states that would be indicative of 
an unwanted Goldstone boson.

The model under discussion is defined by a partition function describing
an ensemble of hyperspherical domains. Each domain is characterised 
by a background covariantly constant self-dual or anti-self-dual
gluon field of random orientation. Summing over all orientations
and dualities guarantees a Lorentz and CP invariant ensemble.
Confinement of  quarks is manifested in the model,
as demonstrated in the original work \cite{NK2001}.
On the boundaries of each hypersphere, fermion fluctuations 
satisfy a chirality violating 
boundary condition
\begin{equation} 
i\!\not\!\eta(x) e^{i\alpha\gamma_5}\psi(x)=\psi(x)
\label{prelimbc}
\end{equation}
which is $2\pi$ periodic in the chiral angle $\alpha$.
Here $\eta_{\mu}$ is a unit radial vector at the boundary.  
Integrating over all such chiral angles guarantees
chiral invariance of the ensemble.  
As a consequence of Eq.~(\ref{prelimbc}) 
the spectrum of eigenvalues $\lambda$ of the Dirac operator
in a single domain is asymmetric under $\lambda\rightarrow -\lambda$.
Such asymmetries have been studied in other contexts,
for example by \cite{DGS98}.  
In the case of the domain model, the above boundary conditions are   
combined with the (anti-~)self-dual gluon field
which leads to a strong correlation between the
local chirality of quark modes at the centres of domains
with the duality of the background gluon field
\cite{NK2002}.
In this paper we study how these aspects contribute
to quark condensate formation and the pattern of chiral symmetry
breaking in the vacuum.

The vacua of the quantum problem associated with an 
ensemble of domains are the minima
of the free energy determined from the partition function.
The problem of the quark contributions to the free energy
requires calculation of the determinant of the Dirac operator 
in the presence of chirality violating boundary conditions.
For slightly different choice of boundary 
condition (with $\alpha\to -i \vartheta-\pi/2$) 
this problem has been addressed
in \cite{WD94} where the parity odd part
of the logarithm of the determinant was identified as 
$\ln {\rm det} (i\!\not\!D) \sim   2q\vartheta$ 
with  $q$ the topological charge (not necessarily integer) 
of the underlying gluon field. 
This is basically the chiral anomaly.
For the specific gluon field relevant to
the domain model we have obtained a similar result 
for the parity odd part \cite{NKBari,NK2003}
\begin{equation}
\ln {\rm det} (i\!\not\!D) \sim 2 i q (\alpha \ {\rm mod} \pi),
\label{prelimanomaly}
\end{equation}
consistent with \cite{WD94}.
We study the consequences of this result in the context of 
the domain model. 

A nonzero quark condensate is generated in the model
without there being a continuous chiral $U_A(1)$ degeneracy of
minima of the ensemble free energy.   
There are two factors 
which interplay to achieve this result. Firstly,
when self-dual and anti-self-dual configurations are summed,
the anomaly Eq.~(\ref{prelimanomaly}) leads to a contribution
to the free energy of the form $-\ln\cos(2q\arctan(\tan\alpha))$ which
vanishes when $\alpha=0,\pi$. The vacuum is degenerate with respect to $Z_2$
chiral transformations. The second factor is the spectral
asymmetry itself which in the presence of an infinitesimally small
quark mass leads to an addition to the free energy linear in the mass,
which removes degeneracy between two discrete minima, 
and thus generates the nonzero quark condensate.
This gives a model with the chiral $Z_2$  discrete subgroup of $U_A(1)$ 
being spontaneously broken, and not the continuous $U_A(1)$ itself.
In the absence of the mass term the ensemble average 
of $\bar\psi \psi$ correctly vanishes. 

Moreover, the form of Eq.(\ref{prelimanomaly})
means that the free energy does not depend on flavour nonsinglet 
chiral angles when more than one massless quark flavours are introduced.  
This allows for the correct degeneracy of vacua with respect to continuous 
$SU(N_f)_L\times SU(N_f)_R$
chiral transformations. This vacuum structure implies 
the existence of Goldstone bosons in the flavour nonsinglet 
pseudoscalar channel but not in the singlet channel, which a study of  
the structure of pseudoscalar correlation functions unveils.

\section{THE DOMAIN MODEL}
For motivation and a detailed description of the model we refer the 
reader to \cite{NK2001}.
The essential definition of the model
is given in terms of the following partition function for
$N\to\infty$ domains of radius $R$
\begin{eqnarray}
{\cal Z} & = & {\cal N}\lim_{V,N\to\infty}
\prod\limits_{i=1}^N
\int\limits_{\Sigma}d\sigma_i
\int_{{\cal F}_\psi^i}{\cal D}\psi^{(i)} {\cal D}\bar \psi^{(i)}
\nonumber \\
&&\times \int_{{\cal F}^i_Q} {\cal D}Q^i 
\delta[D(\breve{\cal B}^{(i)})Q^{(i)}]
\Delta_{\rm FP}[\breve{\cal B}^{(i)},Q^{(i)}]
\nonumber \\
&&\times e^{
- S_{V_i}^{\rm QCD}
\left[Q^{(i)}+{\cal B}^{(i)}
,\psi^{(i)},\bar\psi^{(i)}
\right]}
\label{partf}
\end{eqnarray}
where the functional spaces of integration
${\cal F}^i_Q$ 
and ${\cal F}^i_\psi$  are specified by the boundary conditions  
$(x-z_i)^2=R^2$
\begin{eqnarray}
\label{bcs}
&&\breve n_i Q^{(i)}(x)=0, 
\\
&&i\!\not\!\eta_i(x) e^{i\alpha_i\gamma_5}\psi^{(i)}(x)=\psi^{(i)}(x),
\label{quarkbc} \\
&&\bar \psi^{(i)} e^{i\alpha_i\gamma_5} i\!\not\!\eta_i(x)=-\bar\psi^{(i)}(x).
\label{adjquarkbc} 
\end{eqnarray}
Here $\breve n_i= n_i^a t^a$ with the 
generators $t^a$  of $SU_{\bf c}(3)$ in the adjoint representation
and the $\alpha_i$ are chiral angles 
associated with for the boundary condition
Eq.(\ref{quarkbc}) with different values randomly assigned to domains. 
We shall discuss this constraint in detail in later sections.
The thermodynamic limit assumes $V,N\to\infty$ but 
with the density $v^{-1}=N/V$ taken fixed and finite. The
partition function is formulated in a background field gauge
with respect to the domain mean field, which is approximated 
inside and on the boundaries of the domains by
a covariantly constant (anti-)self-dual gluon field with the 
field-strength tensor of the form
\begin{eqnarray}
F^{a}_{\mu\nu}(x)
&=&
\sum_{j=1}^N n^{(j)a}B^{(j)}_{\mu\nu}\vartheta(1-(x-z_j)^2/R^2), \nonumber \\
B^{(j)}_{\mu\nu}B^{(j)}_{\mu\rho}&=&B^2\delta_{\nu\rho}.
\end{eqnarray}
Here  $z_j^{\mu}$ are the positions of the centres of domains in Euclidean space.

The measure of integration over parameters characterising domains is 
\begin{eqnarray}
\label{measure}
\int\limits_{\Sigma}d\sigma_i\dots & = & \frac{1}{48\pi^2}
\int_V\frac{d^4z_i}{V}
\int_{0}^{2\pi}d\alpha_i
\nonumber\\
&\times&\int\limits_0^{2\pi}d\varphi_i\int_0^\pi d\vartheta_i\sin\theta_i
\int_0^{2\pi} d\xi_i
\nonumber \\
&\times&\sum\limits_{l=0,1,2}^{3,4,5}
\delta(\xi_i-\frac{(2l+1)\pi}{6})
\nonumber\\
&\times&\int_0^\pi d\omega_i\sum\limits_{k=0,1}\delta(\omega_i-\pi k)
\dots ,
\end{eqnarray}
where $(\theta_i,\varphi_i)$ are the spherical angles of the 
chromomagnetic field, $\omega_i$ is the angle between chromoelectric and 
chromomagnetic fields and $\xi_i$ is an angle parametrising the colour 
orientation. 

This partition function describes a statistical system 
of these domain-like structures, of density $v^{-1}$ where 
the volume of a domain is $v=\pi^2R^4/2$, and where each domain is
characterised by a set of internal parameters and
whose internal dynamics are represented by fluctuation fields.
It respects all the symmetries of the QCD Lagrangian, since the statistical 
ensemble is invariant under space-time and colour gauge symmetries. 
For the same reason, if the quarks are massless then the
chiral invariance is respected. 

Thus the model involves only 
two free parameters: the mean field strength $B$ and the 
mean domain radius $R$. 
Within this framework the gluon condensate to lowest order in fluctuations
is $4B^2$ and the topological charge per domain is $q=B^2R^4/16$.
More significantly, an area law is obtained for static quarks. 
Computation of the Wilson loop for a circular contour  of
a large  radius $L\gg R$ gives a string tension $\sigma = B f(\pi B R^2)$
where $f$ is given for colour $SU(2)$ and $SU(3)$ in \cite{NK2001}
and has a purely geometrical meaning in terms of overlaps of hyperspheres. 
Estimations of the values of these quantities are known from 
lattice calculation or phenomenological approaches and can be used to fit  $B$ and $R$.
As described in \cite{NK2001} these parameters are fixed to
be
$\sqrt{B} = 947 {\rm {MeV}}, R=(760 {\rm{MeV}})^{-1} = 0.26 {\rm {fm}}$ 
with the average absolute value of topological charge per domain
turning out to be $q\approx 0.15$ and the density of domains 
$v^{-1}=42{\rm fm}^{-4}$. The topological susceptibility then
turns out to be $\chi \approx (197 {\rm MeV})^4$, comparable to the 
Witten-Veneziano value \cite{largeNc}. This fixing of the parameters
of the model remains unchanged in this investigation of the quark
sector.  The quark condensate
at the origin of a domain where angular dependence drops out was estimated 
in paper \cite{NK2001} with 
a result $-(228 {\rm{MeV}})^3$. 

\section{DIRAC OPERATOR AND SPECTRUM}
 
The eigenvalue problem 
\begin{eqnarray}
\!\not\!D\psi(x)&=&\lambda \psi(x),
\nonumber\\
i\!\not\!\eta(x) e^{i\alpha\gamma_5}\psi(x)&=&\psi(x), \ x^2=R^2
\end{eqnarray}
was studied in \cite{NK2002}.
Dirac matrices are in anti-hermitean representation.
For $\alpha$ assumed to be real a bi-orthogonal basis has to 
be constructed.
Solutions can be labelled
via the Casimirs and eigenvalues 
\begin{eqnarray*}
{\bf K}_1^2 & = & {\bf K}_2^2 \rightarrow 
            \frac{k}{2}(\frac{k}{2} + 1), \  k=0,1,\dots,\infty \\
K^z_{1,2} & \rightarrow & m_{1,2}, 
\nonumber \\
 m_{1,2}&=&-k/2, -k/2+1,\dots,k/2-1,k/2,
\end{eqnarray*}
corresponding to the angular momentum operators
\begin{eqnarray*}
{\bf K}_{1,2} = \frac{1}{2} ({\bf L} \pm {\bf M})
\end{eqnarray*}
with ${\bf L}$ the usual three-dimensional angular momentum
operator and ${\bf M}$ the Euclidean version of the boost operator.
The solutions for the self-dual background field are then 
\begin{eqnarray}
\psi^{-\kappa}_{km_1}=i\!\not\!\eta\chi^{-\kappa}_{km_1}
+\varphi^{-\kappa}_{km_1},
\label{psikappa}
\end{eqnarray}
where $\chi$ and $\varphi$ must both have negative chirality in the
self-dual field and $\kappa$ related to the polarisation of the field 
defined via the projector
\begin{eqnarray}
O_{\kappa} = N_+ \Sigma_{\kappa} + N_- \Sigma_{-\kappa}
\end{eqnarray}
with 
\begin{eqnarray*}
N_{\pm} = \frac{1}{2}(1\pm \hat n/|\hat n|), \
\Sigma_\pm = \frac{1}{2}(1\pm{ \bf\Sigma  B}/B)
\end{eqnarray*}
being respectively separate projectors for colour and spin polarizations.
 Significantly, the negative chirality for $\chi$ and $\varphi$ is
the only choice for which the boundary condition Eq.(\ref{quarkbc})
can be implemented for the self-dual background. 
The explicit form of the spinors $\chi$ and
$\varphi$ can be found in \cite{NK2002}, where it is demonstrated
that the eigenspinor Eq.~(\ref{psikappa}) has definite chirality at the centre
of domain correlated with the duality of the gluon field.
The boundary condition reduces to
\begin{eqnarray}
\label{bc-main}
\chi=- e^{ \mp i\alpha}\varphi ,  \  \bar\chi= \bar\varphi e^{ \mp i\alpha},
\ x^2=R^2,
\end{eqnarray}
where upper (lower) signs correspond to  $\varphi$ and $\chi$
with chirality $\mp1$, which, using the solutions, amounts to
equations for the two possible polarisations, for $\Lambda^{-+}_{k}$:
\begin{eqnarray}
\label{L-+}
&&e^{-i\alpha}M\left(k+2-\Lambda^2,k+2,z_0\right)
-\frac{\sqrt{z_0}}{i\Lambda}
\nonumber \\
&&\times \left[ M\left(k+2-\Lambda^2,k+2,z_0\right)
\right.\nonumber \\
&&\left.-\frac{k+2-\Lambda^2}{k+2}M\left(k+3-\Lambda^2,k+3,z_0\right)
\right] \nonumber \\
&&=0,
\end{eqnarray}
and for $\Lambda^{--}_{k}$:
\begin{eqnarray}
\label{L--}
&&e^{-i\alpha}M\left(-\Lambda^2,k+2,z_0\right)
\nonumber \\
&&+\frac{i\Lambda \sqrt{z_0}}{k+2}
M\left(1-\Lambda^2,k+3,z_0\right)
=0
\end{eqnarray}
where $z_0 = \hat BR^2/2$ and $\Lambda=\lambda/\sqrt{2\hat B}$.
 For the present work
Eqs.(\ref{L-+},\ref{L--}) are the starting point, from
which we see by inspection that 
a discrete spectrum of complex eigenvalues emerges
for which there is
no symmetry of the form $\lambda\rightarrow -\lambda$.
For given
chirality and polarisation and angular momentum $k$,
an infinite set of discrete $\Lambda$ are obtained labelled
by a ``principal quantum number'' $n$.

\section{QUARK DETERMINANT AND FREE ENERGY FOR A SINGLE DOMAIN}
We consider the one-loop contribution of the quarks to the free energy 
density $F(B,R|\alpha)$
of a single (anti-)self-dual domain of  volume $v=\pi^2 R^4/2$ 
\begin{eqnarray}
\exp\left\{-v F(B,R|\alpha)\right\}
&=&{\rm det}_\alpha\left(\frac{i\!\not \!D}{i\!\not\! \partial}\right)
\nonumber  \\
&=&\prod_{\kappa,k,n,m_1}\left(
\frac{\lambda^\kappa_{kn}(B)}{\lambda^\kappa_{kn}(0) }\right) \nonumber \\
&=&\exp\left\{-\zeta'(s)\right\}_{s=0}.
\label{det1}
\end{eqnarray}
The normalization is chosen such that 
$ \lim_{B\rightarrow 0} F(B,R|\alpha)=0.  $
The free energy is then $F=v^{-1}\zeta'(0)$.
In \cite{NKBari,NK2003}, we discuss in more detail the technical
aspects of the computation of this quantity in zeta function
regularisation, starting with a purely imaginary choice for  $\alpha$ 
which provides for a real spectrum, and analytically continuing the 
final result to real values of $\alpha$.
It is known  that $\zeta(s)$  
has two contributions: from $\zeta_{\!\not\!D^2}(s)$,
which is insensitive to the spectral asymmetry, and from 
the spectral function $\eta(s)$ which is odd under 
$\lambda\rightarrow -\lambda$. These two functions
are defined by
\begin{eqnarray}
\zeta_{\!\not\!D^2}(s)&=&\sum_{k,n,\kappa}(k+1)
\left(\frac{\mu^{2s}}{[\lambda^\kappa_{kn}(B)]^{2s}}
\right. \nonumber \\
&&\left. -
\frac{\mu^{2s}}{[\lambda^\kappa_{kn}(0)]^{2s}}
\right)
\label{zeta1}
\\
\eta(s)&=&\mu^s\sum_{k,n,\kappa}(k+1)\left(
\frac{{\rm sgn}(\lambda^\kappa_{kn}(B))}{|\lambda^\kappa_{kn}(B)|^{s}}
\right.\nonumber \\
&&\left. 
-\frac{{\rm sgn}(\lambda^\kappa_{kn}(0))}{|\lambda^\kappa_{kn}(0)|^{s}}
\right).
\label{eta1}
\end{eqnarray}
In terms of these,
the free energy density for a given parameter $\alpha$ is:
\begin{eqnarray}
F=v^{-1}\left(\frac{1}{2}\zeta'_{\!\not\!D^2}(0)
\pm i\frac{\pi}{2}\zeta_{\!\not\!D^2}(0)\mp i\frac{\pi}{2}\eta(0)\right),
\label{zetaprime}
\end{eqnarray}
with $\mu$  -- arbitrary scale.
The final result for  $F$ is complex with the imaginary part of the
form 
\begin{equation}
\label{anomaly}
\Im{F}= \pm 2 q \rm{Arctan}(\tan(\alpha))
\end{equation}
where $q$ is the absolute value of topological charge in a domain.
This charge is not integer here in general but the anomalous term is 
$\pi n$ periodic in $\alpha$.
This is the Abelian anomaly as observed
within the context of bag-like boundary conditions by
\cite{WD94}. 
Its appearance here is in the spirit of
the derivation by Fujikawa \cite{Fuj80}, where the phase appears as
an extra contribution under a chiral transformation on the
fermionic measure of integration. 
However, in our calculations \cite{NKBari,NK2003},
an $\alpha$ dependent real part also appears. To explain
its possible fate some technical details of the
calculation are necessary. The sum over modes $n$ 
is converted into a contour integral via the relation
\begin{equation}
\sum_{\lambda} {1 \over {\lambda^s}} = \frac{1}{2\pi i}
\oint_{\Gamma} \frac{d\xi}{\xi^s} \frac{d}{d\xi}\ln f(\xi)
\label{intrep}
\end{equation}
where the zeroes of $f$ define the eigenvalues. Thus the 
left hand side of the constraint equations Eqs.(\ref{L-+},\ref{L--})
appear in these expressions. One then performs
a Debye-like expansion in $1/k$ of the Kummer functions.
Exchanging this summation with the sum over $k$ one can
read off each order in $1/k$ via an ordinary Riemann function
whose analytic behaviour is known. However in principle
the order of summations does not commute - there can be extra
terms, which are difficult for computation even in the
simplest examples \cite{Elizbook}. 
Taking into account
such contributions is expected to remove the $\alpha$ dependence
from the real part, leaving only the anomaly contribution from the
imaginary part, consistent with \cite{WD94}. Thus for the
purpose of the following investigation, we will assume 
that the anomaly Eq.~(\ref{anomaly}) provides the  entire result 
for the $\alpha$-dependent part
of the free energy of massless fermions in a domain.

The chiral condensate 
is computed via the presence in the free energy of a term linear in an 
infinitesimally small mass, namely \cite{DGS98} 
\begin{eqnarray}
F=F_{m=0} +
i\frac{m}{\mu v}\eta(1).
\label{massivefreeenergy}
\end{eqnarray}
Using the analogue of the representation Eq.(\ref{intrep})
the  summed  contributions
of both polarisations in the self-dual domain
to the
asymmetry function can be written,
\begin{eqnarray}
\label{eta-1-1}
\eta(s)&=&i \mu R e^{i\alpha} \frac{\cos(\pi s/2)}{\pi(1-s)}
\sum_{k=1}^\infty k^{1-s}\frac{k}{k+1}
\nonumber \\
&&\times \left[
1 + M(1,k+2,z)
\right.\nonumber \\
&&\left.
-\frac{z}{k+2}M(1,k+3,-z)
\right].
\end{eqnarray}
We next evaluate the asymptotic behaviour in $k$.
A singular term
as $s\rightarrow 1$ can be extracted, which turns out to
be field (that is, $B$) independent and is canceled by the normalization. 
The final expression for the term linear in mass in the free energy density
for a self-dual domain is
\begin{eqnarray*}
\delta F^{(\rm sd)}=
- e^{i\alpha} m  \prec\bar\psi \psi\succ,
\end{eqnarray*}
where we have used the suggestive notation
\begin{eqnarray}
\label{cond0}
\prec \bar\psi \psi\succ
&=&\frac{1}{\pi^2 R^3}
\sum_{k=1,z=z_1,z_1,z_2}^\infty \frac{k}{k+1}
\nonumber \\
&&\times \left[
M(1,k+2,z)
\right. \nonumber \\
&&\left. -\frac{z}{k+2}M(1,k+3,-z)-1
\right]
\end{eqnarray} 
coming from $\eta(1)$ with the sum over $z$ 
correponding to a color trace.
The  $\alpha$ dependent part of the free energy of a self-dual domain 
for massive quarks is complex with the following
real and imaginary parts:
\begin{eqnarray}
F&=&\Re{F}+i\Im{F} \nonumber \\
\Re{F} &=& - m \cos \alpha \prec \bar\psi \psi\succ \label{realF} \\
\Im{F} &=& 2 \frac{q}{v}\rm{Arctan}(\tan(\alpha)) 
\nonumber \\
&& - m \sin \alpha \prec \bar\psi \psi\succ.
 \label{imagF}
\end{eqnarray}
The free energy of an anti-self-dual domain
is obtained via complex conjugation. 

\section{GROUND STATE AND CONDENSATE}
Under the assumption that only the anomalous term depends on the chiral angle, 
the part of the free energy 
density ${\cal F}$  relevant for the present consideration of
an ensemble of $N\to\infty$ domains
with both self-dual and anti-self-dual configurations takes the form
\begin{eqnarray*}
e^{-vN{\cal F}}&=&{\cal N}\prod_j^N\int_0^{2\pi}d\alpha_j 
\\
&&\times 
\frac{1}{2}
\left[e^{iv\Im F(\alpha_j)}+e^{-iv\Im F(\alpha_j)} \right]
\nonumber \\
&=&{\cal N}\prod_j^N\int_0^{2\pi} d\alpha_j
e^{ \ln(\cos(v\Im F(\alpha_j)))}
\nonumber
\\
&=&{\cal N}\exp\left(N\max_\alpha
\ln(\cos(v\Im F(\alpha)))\right) .
\nonumber
\end{eqnarray*} 
The maxima (minima of the free energy density)
are achieved at $\alpha_1=\dots=\alpha_N=\pi n$.

In the absence of a quark mass, only the anomaly contribution
in the imaginary part, $\Im{F}$, of the free energy of a single domain
appears under the logarithm of the cosine and defines the minima of the free
energy density, $\ln(\cos(v\Im F(\alpha)))=0$.  
Thus for massless quarks there is no
continuous $U_A(1)$ symmetry in the ground states rather
a discrete $Z_2$ chiral symmetry.
The anomaly plays a peculiar role here: selecting out those
chiral angles which minimise the free energy
so that the full $U_A(1)$ group is no longer reflected in
the vacuum degeneracy.
It should be stressed here that this residual discrete degeneracy is sufficient
to ensure zero value for the quark condensate in the absence of mass term
or some other external chirality violating sources.

In a different and more general context the
idea of dynamical relaxation of the effective $\theta$-parameter  
based on an energy minimization criterion was discussed in great
detail in~\cite{Mink}. 

Now switching on the quark mass, we see this discrete
symmetry spontaneously broken, and one of the two vacua selected in the
infinite volume limit according to the sign of the mass.
In this case (for these conventions of boundary
condition and mass term), it is the
minimum at $\alpha=0$. 
The quark condensate can be now extracted from the free energy via
\begin{eqnarray*}
\langle \bar\psi(x)\psi(x)\rangle&=&-\lim_{m\to0}\lim_{N\to\infty} 
(vN)^{-1}\nonumber \\
&&\times \frac{d}{dm}e^{-vN{\cal F}(m)}.
\end{eqnarray*} 
Taking the thermodynamic limit $N\to\infty$ and then $m\to0$ gives  
a nonzero condensate 
\begin{eqnarray*}
\langle \bar \psi(x) \psi(x)\rangle =-\prec \bar \psi \psi\succ.
\end{eqnarray*} 
According to Eq.~(\ref{cond0}) the condensate is equal to 
\begin{eqnarray*}
\langle \bar \psi(x) \psi(x)\rangle  = -(237.8 \ {\rm MeV})^3
\end{eqnarray*} 
for the values of field strength $B$ and domain radius $R$ 
fixed earlier by consideration of the pure gluonic 
characteristics of the vacuum -- string tension, topological succeptibility and 
gluon condensate.
A nonzero condensate is generated without 
a continuous degeneracy of the ground states of the
system.

\section{MULTIFLAVOUR CASE}
The question remains though whether any continuous directions in the vacua
to be expected when the full flavour chiral symmetry is brought
into play. For this we must generalise the analysis. 
We consider $N_f$ massless quark flavours.
Firstly, we observe that the fermion boundary
condition in Eq.(\ref{quarkbc}),
explicitly breaks all chiral symmetries, flavour singlet
and non-singlet (see also \cite{WD94}). 
Thus the procedure we have used here of integrating
over all $\alpha$ does not suffice to restore the full chiral
symmetry of the massless QCD action. Rather, the boundary
condition must be generalised to include flavour non-singlet
angles,
\begin{equation}
\alpha \rightarrow \alpha +\beta^a T^a,
\end{equation}
with $T^a$ the $N_f^2-1$ generators of $SU(N_f)$. 
Then integration over $1+N_f^2-1 =N_f^2$ angles 
$\alpha,\beta^a \in [0,2\pi]$ must be performed
for a fully chiral symmetric ensemble. The spectrum of the Dirac
problem now proceeds quite analogously, except that the
boundary condition mixes flavour components, thus an additional
projection into flavour sectors
is required in order to extract the eigenvalue equation analogous to
Eq.(\ref{bc-main}). 

For $N_f=2$ the projection is easy to write
down, 
$(1 \pm {\hat \beta}\cdot {\vec \sigma})/2$, which will have
the effect that 
 $\alpha$ in the dependence
of eigenvalues and spectral functions is replaced 
by $\alpha \pm |{\vec \beta}|/2$ for the two $N_f=2$ isospin 
projections.  
Thus for a given domain, the imaginary part of
the free energy will
be the sum of two terms involving the imaginary part determined
for a single flavour
but evaluated at the two shifted angles:
\begin{eqnarray}
\Im F(\alpha +|{\vec \beta}|/2)+
\Im F(\alpha -|{\vec \beta}|/2) 
=  \Im F(\alpha )
\label{twoflavourimag}
\end{eqnarray}
since $\Im F(\alpha )=2q \arctan(\tan(\alpha))=2q(\alpha+n \pi)$. 
This expresses the known fact that the anomaly
only appears in the flavour singlet sector or
is Abelian. Thus for an ensemble of domains
the free energy is identical to that for one massless
flavour, namely it depends only on the Abelian angle
$\alpha$. 
Thus for $N_f=2$, 
the $U_A(1)$ direction remains fixed by energy minimisation
while the $SU(2)_L\times SU(2)_R$ directions represent degeneracies
in the space of ground states in 
the thermodynamic limit.

The generalisation to
$SU(N_f)_L\times SU(N_f)_R$ can be summarised as follows:
the functions $\pm |{\vec \beta}|$ in Eq.~(\ref{twoflavourimag}) become 
functions $B^i(\beta^a)$ which correspond to the number
of diagonal generators of the flavour group. 
The cancellation in Eq.(\ref{twoflavourimag}) generalises
reflecting the tracelessness of the $SU(N_f)$ generators,
$\sum_i B^i(\beta^a)=0$.
Thus for $N_f=3$ one expects {\it eight} not nine continuous
directions in the space of vacua.

The  consequence of this realisation of chiral symmetries
in the meson spectrum can be seen in the general structure
of correlators of the
flavour nonsinglet $J^{a}_{\rm P}(x)$ and 
singlet $J_{\rm P}(x)$ pseudoscalar
quark currents as they appear in the domain model,
\begin{eqnarray*}
\langle  J^{a}_{\rm P}(x) J^{b}_{\rm P}(y)\rangle
&=&\overline{\langle \langle J^{a}_{\rm P}(x) J^{b}_{\rm P}(y)\rangle\rangle}
\\
\langle  J_{\rm P}(x) J_{\rm P}(y)\rangle
&=&\overline{\langle \langle J_{\rm P}(x) J_{\rm P}(y)\rangle\rangle}
\\
&-&
\overline{\langle \langle J_{\rm P}(x)\rangle\rangle \langle\langle J_{\rm P}(y)\rangle\rangle}.
\end{eqnarray*}
Here double brackets denote integration
over quantum fluctuation fields and the overline means
integration over all configurations in the domain ensemble.
The second line in the right hand side
of the  flavour singlet correlator reflects the inhomogeneity of the 
background field for a particular domain configuration, translation invariance
being restored only for the ensemble average. 
The second term is entirely determined by the correlation function of the 
background gluon field $B$ in the ensemble~\cite{NK2001}. 
The analogous ``disconnected term'' in the  flavour nonsinglet correlator
is equal to zero due to the trace over flavour indices.
It should be added that the pseudoscalar condensate,
$\langle \bar{\psi} \gamma_5 \psi \rangle$, naturally vanishes 
since parity is not broken in the ensemble of domains.
Thus massless modes can be expected in the  nonsinglet
channel, but not in the flavour singlet due to the 
additional term in the correlator. This general structure of correlators is 
exactly the same as in the 
instanton liquid model~\cite{Shuryak} but manifests
the mechanism for eta-prime mass generation
proposed by Witten in \cite{largeNc} and appreciated in chiral
perturbation theory by \cite{chpert}.

\end{document}